\documentclass{article} 

\usepackage{amsmath,amssymb,amsthm,latexsym} 

\usepackage{stmaryrd,wasysym,upgreek,mathrsfs,dsfont} 
\usepackage[english]{babel} 
\usepackage{graphicx,color} 

\usepackage{array}

\newtheorem{lemma}{Lemma}[section]

\newtheorem{theorem}{Theorem}[section]
\newcommand{\bee}{\begin{equation}}
\newcommand{\ee}{\end{equation}}
\newcommand{\bea}{\begin{eqnarray}}
\newcommand{\eea}{\end{eqnarray}}

\newcommand{\Tr}{{\rm Tr}}

\newcommand{\cT}{{\cal{T}}}

\newcommand{\cF}{{\cal{F}}}

\newcommand{\prf}{{\noindent {\rm \bf Proof}\, }}

\begin{document} 

\title{Loop Vertex Expansion\\ for 
Higher Order Interactions}
\author{Vincent Rivasseau\\
Laboratoire de Physique Th\'eorique, CNRS UMR 8627,\\ 
Universit\'e Paris XI,  F-91405 Orsay Cedex, France}

\maketitle 
\begin{abstract} 
This note provides an extension of the constructive \emph{loop vertex expansion}
to stable interactions of arbitrarily high order, opening the way
to many applications. We treat in detail the example of the $(\bar \phi \phi)^p$ 
field theory in zero dimension. We find that the important feature to extend 
the loop vertex expansion is not to use an intermediate field representation, but rather to force 
integration of exactly one particular field per vertex of the initial action. 
\end{abstract} 

\begin{flushright}
LPT-20XX-xx
\end{flushright}
\medskip

\noindent  MSC: 81T08, Pacs numbers: 11.10.Cd, 11.10.Ef\\
\noindent  Key words: Constructive field theory, Loop vertex expansion.

\medskip

\section{Introduction}

The loop vertex expansion (LVE) \cite{Rivasseau:2007fr} 
combined an intermediate field representation
with a replica trick and a forest formula \cite{BK,AR1} 
to express the cumulants of a Bosonic field theory with quartic interaction 
in terms of a convergent sum over trees. This method has many advantages:

\begin{itemize}

\medskip\item 
like Feynman's perturbative expansion, it allows to compute connected quantities at a glance: the partition function
of the theory is expressed by a sum over forests and its logarithm is exactly the same sum 
but restricted to \emph{connected} forests, i.e. trees,

\medskip\item 
the functional integrands associated to each forest or tree are absolutely and \emph{uniformly} convergent 
for any value of the fields. In other words there is no need for any additional small field/large field analysis,

\medskip\item 
the convergence of the LVE implies Borel summability of the usual perturbation series and the LVE 
directly computes the Borel sum,

\medskip\item  the LVE explicitly repacks infinite subsets of pieces of Feynman amplitudes
to create a convergent rather than divergent expansion for 
this Borel sum \cite{Rivasseau:2013ova}. Such an \emph{explicit} repacking was long thought close to impossible,

\medskip\item 
in the case of combinatorial field theories of the matrix and tensor type \cite{Gurau:2011xp,Guraubook}, suitably rescaled 
to have a non-trivial $N \to \infty$ limit \cite{Hooft,Gurau:2010ba,Gurau:2011aq,Gurau:2011xq}, the
Borel summability obtained in this way is \emph{uniform} in the size $N$ of the model 
\cite{Rivasseau:2007fr,Gurau:2014lua,Gurau:2013pca,Delepouve:2014bma}.
We do not know of any other method which can provide yet this type of result,

\medskip\item 
the method can be further developed into a multiscale version (MLVE) \cite{Gurau:2013oqa} to include renormalization 
\cite{Delepouve:2014hfa,Lahoche:2015zya}
\footnote{Here it is fair to add that the models built so far are only of the superrenormalizable type.
Moreover the MLVE is especially adapted to resum the renormalized series
of non-local field theories of the matrix or tensorial type. For ordinary local field theories,
until now, and in contrast with  the more traditional
constructive methods such as cluster and Mayer expansions, it does not conveniently provides
the spatial decay of truncated functions. See however \cite{MR1,Rivasseau:2014bya}.}.

\end{itemize}

For all these reasons it would be nice to generalize the LVE to interactions of order higher than 4, but progress in this direction has been slow. The first attempts were based on oscillating Gaussian integral representations \cite{Rivasseau:2010ke,LR1,Lionni:2016ush}. However these
representations are unsuited for taking absolute values in the integrand.

In this note we propose what we think is the correct extension of the LVE 
to stable Bosonic field theories with polynomial interactions of arbitrarily large order.
We focus on a particular simple example, the $(\bar \phi \phi)^p$ zero dimensional scalar theory, since it 
contains the core of the problem. We derive for this theory a new representation which we call 
the loop vertex representation. The corresponding action is indeed a sum over single loops of arbitrary
order decorated by trees. It is closely related to the generating function 
of the cumulants in the \emph{Gallavotti} field-theoretic representation
of classical dynamical systems \cite{Gal}, and it can be
explicitly written in terms of the Fuss-Catalan \cite{FussCatalan} generating function of order $p$. 
Notice however that such functions
cannot be expressed in terms of radicals of the initial fields for $p > 4$. 
Nevertheless Fuss-Catalan functions are shown rather easily to have bounded derivatives of all orders
(see Theorem \ref{mainthe} below). This is the essential
feature which allows the LVE to work.

Fuss-Catalan functions of order $p$ also govern the leading term in the $N \to \infty$ 
limit of random tensor models of rank $p$ \cite{Bonzom:2011zz}. 
Such models were introduced for a completely different reason, namely to
perform sums over random geometries in dimension $p$ pondered with a discretized form
of the Einstein-Hilbert action \cite{Gurau:2011xp,Guraubook}. This fast-developing approach to quantum gravity 
has been nicknamed the ``tensor track"
\cite{Rivasseau:2011hm}-\cite{Rivasseau:2016zco}.
It attracted further interest recently, when the same models but with 
an additional time dependence were shown to provide the simplest solvable examples
of  quantum holography \cite{Witten:2016iux}-\cite{Bonzom:2017pqs}.
It would be fascinating to better understand why the correct constructive repacking of
Feynman's series for the simplest stable \emph{scalar interactions in zero space-time dimension}
precisely involves the same mathematical functions 
than the simplest models of quantum gravity. 

The first step in this direction 
should be to extend the method presented here to arbitrarily 
high matrix and tensor interactions.
We believe in particular that the
uniform analyticity domains of \cite{Rivasseau:2007fr} and \cite{Gurau:2014lua} 
for the $\Tr M^\dagger M M^\dagger M$ matrix models could be 
extended in this way to matrix models with single trace interaction of arbitrary even order.

Another promising research direction opened by this paper is the construction
of renormalizable matrix and tensor \emph{field theories}
with more complicated propagators 
and stable interactions of degrees higher than 4.
Remark indeed that many interesting tensor field theories use order 6 interactions \cite{BenGeloun:2011rc,Carrozza:2013wda}
and cannot be treated therefore with the ordinary quartic LVE. 

Although we treat only complex fields in this paper for simplicity, we think that with relatively minor modifications
our method can be extended to real-field models with typical interactions of the 
$\phi^{2p}$ type instead of $(\bar\phi\phi)^{p}$. The key idea should be to
force again integration of a single field per vertex. This generates 
rooted trees with a single external face in addition to the single-loop diagrams
with two faces of Fig. \ref{barb1}-\ref{barb2}.
 
\section{Loop Vertex Representation} \label{lvr}
\medskip
Let us fix an integer $p \ge 2$.
The $(\bar \phi \phi)^p$ model is defined by the partition function with sources
\begin{equation}
Z_{p}(\lambda, \bar J, J)  = \int d \mu (\phi , \bar \phi ) e^{-\lambda (\bar\phi \phi)^{p}+ \bar J \phi + J \bar \phi}  ,
\end{equation}
where $d \mu (\phi , \bar \phi )$ is the Gaussian normalized measure o covariance 1 for th epair of complex conjugate fields 
$\phi$ and $\bar \phi$. Hence  $ \int d \mu (\phi , \bar \phi )   \phi^r \bar \phi^s  = \delta_{rs} r! $.
The (divergent) perturbative power series in $\lambda$ writes
\bea
Z_{p}(\lambda, \bar J, J) &=& 
\sum_{q,n =0}^\infty  (-\lambda)^n  \frac{(pn +q)!}{n!} \frac{(\bar J J)^q}{(q!)^2} .
\eea

We write simply $Z_{p}(\lambda)$ for the normalization of the theory:
\begin{equation}
Z_{p}(\lambda)  = \int d \mu (\phi , \bar \phi ) e^{-\lambda (\bar\phi \phi)^{p}} =Z_{p}(\lambda, \bar J, J)\vert_{J = \bar J =0} .
\end{equation}

The $2N$-th connected moments (or $2N$th cumulants) are given by
\bea  G^c_{p,N} (\lambda) :=  \bigl[\frac{\partial}{\partial J }\frac{\partial}{\partial \bar J }\bigr]^N \log
Z_{p}(\lambda, \bar J, J)  \vert_{J = \bar J =0} .
\eea
A main goal in field theory is therefore to compute the logarithm of
\bea
Z_{p} (\lambda, \bar J , J) 
&=&   \sum_{q=0}^\infty \int  d \mu (\phi , \bar \phi ) \frac{(J\bar J)^q}{(q!)^2}\sum_{n=0}^\infty 
\frac{(pn+q)!}{((p-1)n)!n!} [-\lambda (\bar \phi \phi)^{p-1}]^n\nonumber\\
&=& \sum_{q=0}^\infty \int  d \mu (\phi , \bar \phi ) \frac{1}{(q!)^2} 
[J\bar J  \frac{\partial}{\partial g} ]^q  \sum_{n=0}^\infty g^{pn+q}   {{pn}\choose{n}}  ( \bar \phi \phi)^{(p-1)n} \nonumber\\
&=&   \sum_{q=0}^\infty \int  d \mu (\phi , \bar \phi ) \frac{1}{(q!)^2} 
[J\bar J  \frac{\partial}{\partial g} ]^q  g^q  e^{S_p(g, \phi , \bar \phi)}    \label{deflvr}
\eea
where in the second line we define $ g = (-\lambda)^{\frac{1}{p}}$ and for the third line we define
\bea
F_{p}(g, \phi , \bar \phi) &=& \sum_{n=0}^\infty   {{pn}\choose{n}} g^{pn} (\bar \phi \phi)^{(p-1)n},\\
S_p(g, \phi , \bar \phi) &=&  \log F_{p}(g, \phi , \bar \phi).
\label{def1}
\eea
We call  \eqref{deflvr} the loop vertex representation (LVR) of the theory\footnote{This terminology 
follows from the graphical representation of $S_p$ given in  Section \ref{graphicint}.}.
In this LVR the normalization $Z_p$ is given by
\bee  Z_p(\lambda) = \int  d \mu (\phi , \bar \phi )  e^{S_p} 
\ee
and the connected 2-point function $G^c_{p,1} (\lambda) $ 
is the same as the normalized 2 point function, hence its LVR representation is
\bea  G^c_{p,1} (\lambda) &=& \frac{1}{Z_p(\lambda)} \frac{\partial}{\partial J }\frac{\partial}{\partial \bar J }
Z_{p}(\lambda, \bar J, J)  \vert_{J = \bar J =0}\\
&=& \frac{1}{Z_p(\lambda)} \int  d \mu (\phi , \bar \phi )  \frac{\partial}{\partial g}  g e^{S_p}  \\
&=& 1  + \frac{g}{Z_p(\lambda)}\int  d \mu (\phi , \bar \phi )  \frac{\partial  S_p}{\partial g}  e^{S_p},
\eea
Remark that the term 1 corresponds to the free two point function.

$F_p$ and $S_p$ are solely functions of $z= g^p (\bar \phi \phi)^{p-1}  = -\lambda (\bar \phi \phi)^{p-1}$, which will be also
denoted $F_p$ and $S_p$ through some abuse of notations. More precisely
\bea
S_p(z) =  \log F_{p}(z) , \quad
F_{p}(z) = \sum_{n=0}^\infty   {{pn}\choose{n}} z^n.\label{defFS}
\eea

The binomial coefficient
${{pn}\choose{n}}= \frac{(pn)!}{n!((p-1)n)!}$ is not far from the  $p$th Fuss-Catalan number \cite{FussCatalan}
$C_{n}^{(p)} := \frac{1}{pn+1 }{{pn+1}\choose{n}} = 
\frac{1}{(p-1)n+1 }{{pn}\choose{n}}$.
We know that the generating function
\begin{equation}
T_p(z) = \sum_{n=0}^\infty C_{n}^{(p)} z^n \label{gencat1}
\end{equation}
for such generalized Fuss-Catalan numbers obeys the algebraic equation 
\begin{equation}\label{gencatalan}
zT_p^p(z) -T_p(z) +1 =0 .
\end{equation}
It governs also the enumeration of melonic graphs at rank $p$ \cite{Bonzom:2011zz}.
Equation \eqref{gencatalan} is soluble by radicals for $p\le 4$ but not beyond,
for  $p>4$ \cite{Perrin}-\cite{Osada}.

By deriving  this equation we find 
\bea
F_{p}(z) &=& \sum_{n=0}^\infty   {{pn}\choose{n}} z^n = (p-1) z T_p' +T_p \\
&=& \frac{T_p}{p - (p-1)T_p}  =  \frac{1}{1 - p z  T_p^{p-1}}.  \label{f0}
\eea
Therefore the action $S_p$  computes explicitly in terms of $T_p$ as
\bea
S_p &=& - \log [1 - p z  T_p^{p-1}] = \sum_{q=1}^\infty  \frac{1}{q} [p z  T_p^{p-1}]^q .
\eea
In the simple case $p=2$ we know that $T_2 (z) = \frac{1 - \sqrt{1 - 4z}}{2z}$, hence
\bee
S_2 = - \frac{1}{2}\log [1 - 4 z ] = \sum_{q=1}^\infty  \frac{1}{2q} (4z)^q .
\ee

The loop vertex expansion (LVE) rewrites
\bea   \label{lve1}
Z_{p} (\lambda, \bar J , J)
&=&   \sum_{q=0}^\infty \int  d \mu (\phi , \bar \phi )  \frac{1}{(q!)^2} 
[J\bar J  \frac{\partial}{\partial g} ]^q  g^q \sum_{n=0}^\infty \frac{S_p^n}{n!}   
\eea
and after applying $q$ derivatives $ \frac{\partial}{\partial g}$ which either derive the trivial
$g^q$ factor or derive a certain number of ``marked" loop vertices $S_p$,
it applies an interpolation formula between replicas of the fields in all (marked or unmarked) vertices to write $Z_{p} (\lambda, \bar J , J)$
as a sum over \emph{forests}.
$\log Z_{p} (\lambda, \bar J , J)$ is then given simply by \emph{exactly the same sum} but restricted 
to connected forests i.e. \emph{trees}. Convergence of
the LVE depends on good bounds on the derivatives of $S_p$. Our next section addresses this question.

\section{Properties of $F_p$ and $S_p$}


Consider the function $S= -\log (1 - z) = \sum_{n\ge 1} \frac{z^n}{n}$. 
It is well defined in the cut-plane $\mathbb{C}^{cut}:= \mathbb{C} - [1 + \infty]$, 
it is not bounded in that domain but its derivative of order $q$ is $S^{(q)} = (q-1)!(1-z)^{-q}$,
which is bounded in modulus by $(q-1)!  [\frac{K(\epsilon)}{1 + \vert z \vert }]^q $
if we exclude a sector of small opening angle $\epsilon $ around the positive real axis.
These are in fact exactly the properties which allow the LVE to work. The action $S_p$
is not as simple, but has exactly the same properties, provided we replace 1 by the convergence
radius $R_p = \frac{(p-1)^{p-1}}{p^p}$ of the Fuss-Catalan functions.
More precisely

\begin{theorem}\label{mainthe}
$T_p$, $F_p$  and $S_p$ are analytic functions of $z$ 
in the cut plane $\mathbb{C}^{cut}_{p}:= \mathbb{C} - [R_p, + \infty]$ where $R_p = \frac{(p-1)^{p-1}}{p^p}$.
For any $\epsilon >0 $, in the open sector  $\mathbb{C}_{\epsilon} := \mathbb{C} -\{ z , \vert  \arg z \vert  \le \epsilon \}$, $S_p$
can grow only logarithmically when $\vert z \vert \to \infty$, and there exists a constant $K_p ( \epsilon) >0$ 
such that for any $q>0$ the $q$-th derivative of $S_p$ is bounded by
\bee \vert S_p^{(q)}  (z)\vert \le (q-1)! \bigl[ \frac{K_p ( \epsilon)}{1 + \vert z \vert }\bigr]^q   \label{goodbound}.
\ee 
\end{theorem}
\prf  We proceed in steps and some intermediate lemmas will occur along the argument.
Let us fix the integer $p$ and write $T$, $F$, $S$ \dots for $T_p$, $F_p$, $S_p$ \dots , 
$T'$, $F'$ \dots for  $\frac{d T}{dz}$, $\frac{dF}{dz}$ \dots and $T^{(q)}$, $F^{(q)}$ \dots for  $\frac{d^q T}{dz^q}$, $\frac{d^qF}{dz^q}$ \dots. 

By Stirling's formula,  $T$ is analytic in the disk $D_p= \{z , \vert z \vert < R_p\}$. 
Clearly in its maximal domain of analyticity $D_{max}$, the functional equations
\bea\label{gencatalan1}
&&zT^p  - T +1 =0 , \\ 
&&T' = \frac{T^p}{1 - pz T^{p-1}} = \frac{T(T-1)}{z[(p-1)T -p]}\label{gencatalan2}
\eea
hold. \eqref{gencatalan1} implies that $T$ is uniformly bounded away from 0 in any compact of $D_{max}$. For 
large $z$ in $D_{max}$, the equation also implies that $T$ must tend to zero as $(-z)^{-1/p}$ so that $z T^p$ tends to $-1$. 
Knowing that and using the implicit function theorem\footnote{We thank A. Sokal for pointing this argument to us.}, \eqref{gencatalan2}
implies that $T$ can have a singularity  
only at points where both $(p-1)T=p$, hence $T= \frac{p}{p-1}$ and $z= \frac{1}{pT^{p-1}} = \frac{(p-1)^{p-1}}{p^p} = R_p$.
(hence clearly by Pringsheim theorem on power series with positive coefficients
$\lim_{z \to R_p } T(z) = \frac{p}{p-1}$).
Therefore $D_{max}$ includes at least the cutplane  $\mathbb{C}^{cut}_{p}$. 
Since  
\bee  F 
= \frac{T}{p - (p-1)T}  
=  \frac{1}{1 - p z  T^{p-1}},
\ee
$F$ is also analytic in $D_{max}$, and it is bounded uniformly away from 0
in all of $D_{max}$. In particular it cannot vanish and $S = \log F$ is 
therefore also analytic in $D_{max}$.

Our next step is to prove a uniform decay of $F$ at infinity in the  open sector  $\mathbb{C}_{\epsilon} $.
More precisely if we define $E:= FT^{p-1}$
\begin{lemma} 
For $z \in \mathbb{C}_{\epsilon} $ we have
\bea \vert  F(z) \vert &= \vert  (1 - pz T^{p-1})^{-1} \vert 
&\le \frac{K_p (\epsilon)}{(1 + \vert z\vert)^{1/p}} \label{decay1}\\
 \vert  E(z) \vert &= \vert F(z)T^{p-1}(z) \vert &\le \frac{K_p (\epsilon)}{1 + \vert z\vert} \label{decay2}
\eea 
for some constant $K_p ( \epsilon)$.
\end{lemma}

\prf We know that $\vert T \vert $ can tend to 0
only for $\vert z \vert  $ large enough, say for $z$ in
$\mathbb{C}_{\epsilon}^{> K} = \mathbb{C}_{\epsilon} \cap \{z,  \vert z \vert >K\}$ and that it indeed tends to 0 as $(-z)^{-1/p}$ at large $z$.
Therefore the function $\vert pz T^{p-1}  \vert  $ in that case tends to $+\infty$ at least as
$p \vert z\vert^{1/p } -1$, hence $\vert  (1 - pz T^{p-1})^{-1} \vert  \le K_p (1 + \vert z \vert )^{-1/p}$ for some constant $K_p$.
In the complement $\mathbb{C}_{\epsilon}^{\le K} = \mathbb{C}_{\epsilon} \cap \{z,  \vert z \vert \le K\}$, whose closure is compact, 
$z$ remains bounded away from the only singularity $z = R_p$ where $1 - pz T^{p-1}$ vanishes, hence
$\vert  (1 - pz T^{p-1})^{-1} \vert $
is bounded by a constant (which depends on $\epsilon)$. 
We conclude that \eqref{decay1} holds for some constant $K_p ( \epsilon)$ in the whole sector $\mathbb{C}_{\epsilon}$
and \eqref{decay2} follows by a similar argument since $\vert T\vert $ also  decays at infinity as  $(1 + \vert z \vert )^{-1/p}$.
\qed

In particular the Lemma implies that $S$ can grow to infinity only logarithmically when $\vert z \vert \to \infty$. 
The next step is to compute and bound the derivatives of $S$. We first remark that 
\bee T' = ET, \quad  \frac{F'}{F} = pE (1 + (p-1) z E).
\ee
From this we can deduce the successive derivatives of $E$:
\bea 
E' &=& E^2 [ (2p-1) + p(p-1)zE] \\
E" &=&  (7p^2 - 9p +2) E^3  + 2p(p-1)(5p-2) zE^4 + 3 p^2(p-1)^2 z^2 E^5 \cdots \nonumber
\eea
and prove easily by induction that 
$E^{(q)}$ is a sum of at most $K^q q!$ monomials of the type $E^{q+1}$, $zE^{q+2}$, \dots $z^{q}E^{2q-1}$.
Hence for $q >0$
\bee S^{(q)} =  [\frac{F'}{F} ]^{(q-1)} = [pE (1 + (p-1) z E)]^{(q-1)}
\ee
is also a sum of at most $K^q q!$ monomials of the type $E^{q}$, $zE^{q+1}$, \dots $z^{q-1}E^{2q-1}$.
This together with \eqref{decay2} completes the proof of \eqref{goodbound} hence of Theorem \ref{mainthe}. \qed

\section{Graphical Representation} \label{graphicint}
Let us give another equivalent form of $S_p$ which allows for a clearer graphical interpretation.

\begin{figure}[!h]
\begin{center}
{\includegraphics[width=7cm]{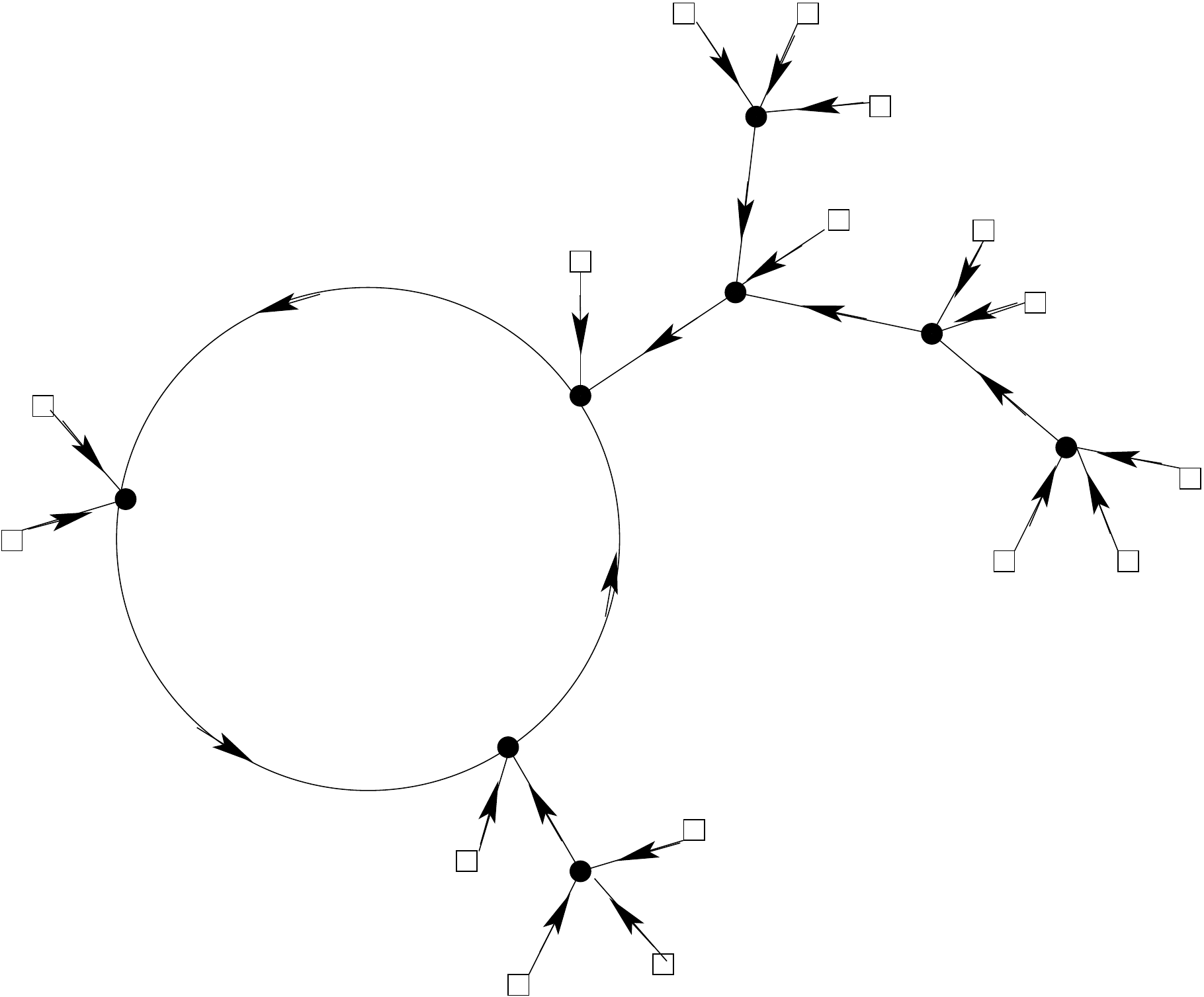}}
\end{center}
\caption{A connected graph of the Gallavotti theory. Black dots are vertices and bear $\lambda$ factors, white squares bear $J$ factors and arrows point from $\bar\phi$
to $\phi$. Remark that three arrows point to each vertex so this is a connected graph of the $\lambda\bar\phi \phi^{3} + \bar\phi J$ theory.}
\label{barb1}
\end{figure}

Consider the Gallavotti theory \cite{Gal} with partition function
\begin{equation}
Z^G_{p}(\lambda, J)  = \int d \mu (\phi , \bar \phi ) e^{\lambda \bar\phi \phi^{p}+ J \bar \phi} .
\end{equation}
Expanding the exponential we get
\begin{equation}
Z^G_{p}(\lambda, J)  = \sum_{n=0}^\infty  {{pn}\choose{n}} \lambda^{n} J^{(p-1)n}  = F_p (z),
\end{equation}
with $z := \lambda J^{p-1}$.
Graphically if we orient edges in the direction $\phi$ to $\bar \phi$ this series represents the sum
over arbitrarily many oriented cycles of arbitrary length decorated with \emph{oriented regular p-ary trees} pointing towards the cycles (see Figure \ref{barb1} for an example with $p=3$). The weights correspond to a $J$ factor at every leaf and a $\lambda$ factor at every node.

We can therefore identify the free energy ${\cal A}_p(\lambda , J) =\log Z^G_p (\lambda , J)$ of that theory 
to the same sum restricted to connected graphs, hence to a sum over \emph{single oriented cycles of arbitrary length
$q$} (with regular associated cycle weight $\frac{1}{q}$) \emph{decorated with oriented regular p-ary trees} pointing towards the cycles. We therefore understand that the computations of section \ref{lvr} correspond to force integration over a \emph{single 
$\bar \phi$ field per vertex}, keeping all others as frozen spectators.

\begin{figure}[!h]
\begin{center}
{\includegraphics[width=7cm]{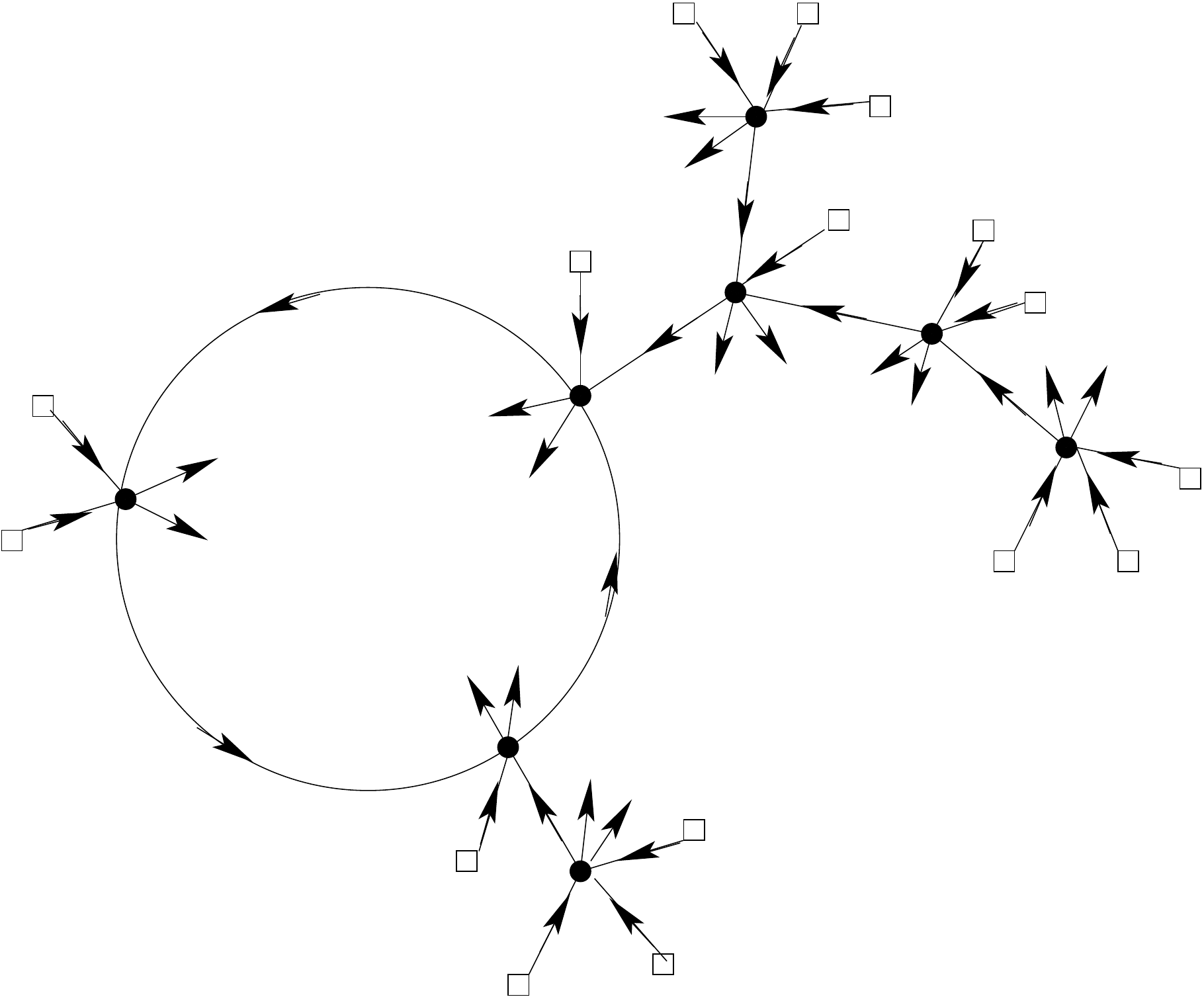}}
\end{center}
\caption{The same connected graph but reinterpreted for the $(\bar\phi \phi)^3$ theory.
White squares are now $\phi$ factors, outgoing arrows are factors $\bar \phi$. Remark
that each vertex is now 6-valent.}
\label{barb2}
\end{figure}

A graphical representation of the action of the previous section can be deduced by changing the sign of $\lambda$,
substituting factors $\phi$ for each $J$ factor and adding a factor $\bar \phi^{p-1}$ at each node, see Figure \ref{barb2}. In other words
\bee  S_p(g, \phi , \bar \phi) = {\cal A}_p(\lambda , J)\vert_{J= \phi, \lambda = g^p  \bar \phi^{p-1}} .
\ee

We think these figures show also convincingly how the method extends to other $(\bar\phi\phi)^{p}$-type theories
with more complicated multidimensional Gaussian measures having less trivial propagators $\Gamma$, such as those required
by usual $d$-dimensional field theories with inverse Laplacian propagator or by matrix and tensor models and field theories. 
Simply add these propagators on the edges of the graphs in Figures \ref{barb1}-\ref{barb2}. Such extensions will be studied in future publications. Of course we can also perform such computations for Fermionic theories with Berezin variables, 
but the constructive theory in that case is simpler since sign cancellations make
the perturbative expansion directly summable.

Remark that in the scalar case arrows can be hooked in any way at the vertices
but in the case of $N$ by $N$ matrix models \cite{Rivasseau:2007fr} and \cite{Gurau:2014lua}
or tensor models of rank $r$ with $N^r$ coefficients, edges should be stranded 
and cyclic alternation of arrows at each vertex or insertions on strands of the correct color 
\cite{Guraubook,Gurau:2013pca} should be respected. This will be crucial to ensure analytic estimates 
with the correct scaling in $N$.

\section{Loop Vertex Expansion}

We can then set up the exact analog of the loop vertex expansion for this model.
For simplicity let us compute only $\log Z$. Starting from \eqref{lve1} and applying the LVE
gives, in the notations of \cite{Rivasseau:2016rgt}:
\bea   \label{lve2}
Z_{p} (\lambda)
&=& \sum_n  \int  d \mu (\phi , \bar \phi )   \frac{S_p^n}{n!}   \\
&=& \sum_n \frac{1}{n!}  \sum_{\cF} \int dw_\cF   \int  d\mu_{\cF, w}  (\phi , \bar \phi )  \partial_\cF \prod_{i=1}^n S_p (\lambda, \phi_i,\bar \phi_i) 
\eea
where
\begin{itemize}
\item the sum over $\cF$ is over oriented forests over $n$ labeled vertices $v = 1, \cdots , n$, including the empty forest with no edge. Such forests are exactly the acyclic oriented edge-subgraphs of the complete graph $K_n$.

\item  $\int dw_\cF$ means integration from 0 to 1 over one parameter for 
each forest edge: $\int dw_\cF  \equiv \prod_{\ell\in \cF}  \int_0^1 dw_\ell  $.
There is no integration for the empty forest since by convention an empty product is 1. A generic integration point $w_\cF$
is therefore made of $\vert \cF \vert$ parameters $w_\ell \in [0,1]$, one for each $\ell \in \cF$.

\item  $ \partial_\cF = \prod_{\ell\in \cF} \frac{\partial}{\partial \bar \phi_{i (\ell)}  }\frac{\partial}{\partial \phi_{j (\ell)}} $ means a product of first order partial derivatives with respect to 
the variables $\bar \phi_{i(\ell)}$ and  $\phi_{j(\ell)}$ corresponding to the departure vertex $i(\ell)$ and arrival 
vertex $j(\ell)$ of the oriented line $\ell \in \cF$. Again there is no such derivatives for the empty forest since by convention an empty product is 1.

\item $d\mu_{\cF, w}  (\phi , \bar \phi ) $ is the Gaussian measure on the replica variables $(\phi_i,\bar \phi_i) $
for $i \in \{ 1 , \cdots , n\}$ with covariance $X^\cF (w_\cF)$, which for $i \ne j$,
is the infimum of the $w_\ell$ parameters for $\ell$
in the unique path $P^\cF_{i \to j}$ from $i$ to $j$ in $\cF$. 
If no such path exists, hence $i$ and $j$ belong to different connected components of the forest 
$\cF$, then by convention $X^\cF_{ij} (w_\cF) = 0$. Finally for all $i$ $X^\cF_{ii} (w_\cF ):= 1$.

\end{itemize}

Remember that the symmetric $n$ by $n$ matrix $X^\cF (w_\cF)$ defined in this way is positive for any value of $w_\cF$
so that this formula is well-defined.

Then the formula factorizes over the trees which are the connected components of $\cF$ so that
\bea   \label{lve3}
\log Z_{p} (\lambda)
&=&   \sum_n \frac{1}{n!}  \sum_{\cT} \int dw_\cT   \int  d\mu_{\cT, w}  (\phi , \bar \phi )  \partial_\cT \prod_{i=1}^n S_p (z_i) 
\eea
where
the sum over $\cT$ runs now only over \emph{spanning trees} over the $n$ labeled vertices $i = 1, \cdots , n$,
and $z_i= - \lambda(\phi_i\bar \phi_i)^{p-1}$ where $(\phi_i, \bar \phi_i)$ are the replica variables at vertex $i$.

\begin{theorem} \label{pacman} For any $\epsilon>0$ there exists $\eta$ small enough such that
the sum \eqref{lve3} is absolutely convergent in the ``pacman domain" 
\bee
P(\epsilon, \eta) := \{\lambda \in D(0, \eta), \vert \arg \lambda \vert < \pi - \epsilon\}.
\ee
\end{theorem}
\prf Using Theorem \ref{mainthe}, this reduces to a simple exercise in combinatorics. Indeed in bounding the series
we have just to take into account that the 2$\vert \cT \vert$
derivatives associated to the tree corners will create local factorials $(d_i-1) !$ in the degree of the tree at vertex $i$. 
At each vertex of coordination $d_i$ in the tree
the $d_i$ derivatives with respect to the $\phi_i$ or $\bar \phi_i$
variables in $\partial_\cT$ create indeed a sum of at most  $K^{d_i}(d_i-1) !$ monomials of the type $(-\lambda)^{r_i} \phi^{s_i} \bar \phi^{t_i}  S^{(r_i)}(z_i)$ with
$\sup\{1, \frac{d_i}{2p-2}\} \le r_i \le d_i$ and $s_i + t_i = (2p-3)r_i - (d_i -r_i) = (2p-2)r_i - d_i$
But using \eqref{goodbound}  this sum is therefore bounded by
\bee (d_i -1)! K^{d_i} \bigl[\frac{K_p(\epsilon)}{1 + \vert z_i \vert }\bigr]^{r_i}  \vert z_i\vert ^{r_i  - \frac{d_i}{2p-2}}  \vert \lambda \vert^{\frac{d_i}{2p-2}} 
\le    (d_i -1)! [KK_p(\epsilon)]^{d_i}  \vert \lambda \vert^{\frac{d_i}{2p-2}} .
\ee
Using Cayley's formula for the number of trees with fixed degrees, and taking $\eta$ (hence $\vert \lambda \vert $) in Theorem \ref{pacman} small enough achieves the proof. Notice however that as usual, the case $\cT = \cT_0$, the  ``empty" tree 
reduced to a single loop vertex, 
requires a special treatment. Indeed $S$ itself, in contrast with its derivatives, is unbounded at large $z$, but $S(0) =0$. Hence
we need to write first
$S = \int_0^1 z S' (tz) dt $ and integrate by parts
\bee  \int  d\mu_{\cT_0, w}  (\phi , \bar \phi ) S(z) = -\lambda \int_0^1 dt \int  d\mu_{\cT_0, w}  (\phi , \bar \phi ) \frac{\partial^{p-1}}{\partial \phi^{p-1}} \bigl[\phi^{p-1}  S'(tz) \bigr]
\ee
before applying the previous bounds.
\qed

The convergence of the LVE for the cumulants of the theory essentially amounts to add a finite number of extra 
$\frac{\partial}{\partial g}$ derivatives as \emph{cilia} \cite{Gurau:2013pca} decorating the previous computation. 
It is left to the reader to check that these cilia do not spoil the convergence of the expansion.

\medskip\noindent
{\bf Acknowledgments} We thank G. Duchamp, R. Gurau, L. Lionni and A. Sokal for useful discussions.

\section{Appendix: Explicit Formulas for $p$ small}
\subsection{The $(\bar \phi \phi)^2$ theory}
\medskip
In this case $g= (-\lambda)^{\frac{1}{2}} = i \sqrt \lambda$, $\lambda = - g^2$ and $z = g^2 \bar \phi \phi  = - \lambda \bar \phi \phi $. 
Equation \eqref{gencatalan} takes the form
\bee
zT_2^2(z) -T_2(z) +1 =0 
\ee
with solution the ordinary Catalan function
\bee  T_2(z) = \frac{1- \sqrt {1 -4z}}{2z}.
\ee
We find
\bee
F_2(z) = z T_2' +T_2 =  (1- 4z)^{-1/2} =  (1- 4g^2 \bar \phi \phi)^{-1/2} = (1 + 4 \lambda \bar \phi \phi)^{-1/2} ,
\ee
and the LVR action is 
\bee  S_2 = - \frac{1}{2}\log  (1 -4z) =  - \frac{1}{2}\log  (1- 4g^2 \bar \phi \phi) = - \frac{1}{2}\log (1+4 \lambda \bar \phi \phi).
\ee 
Its first order derivative is
\bee \frac{\partial  S_2}{\partial g}=  \frac{4g \bar \phi \phi }{1- 4g^2 \bar \phi \phi}
\ee 
so that the loop vertex representation of the partition function is
\bea
Z_2(\lambda) &=&   \int  d\mu( \phi, \bar \phi) e^{  - \frac{1}{2}\log (1-4 \lambda \bar \phi \phi)  } ,\\
G^c_{2,1} (\lambda) &=& 1 - \frac{\lambda}{Z_2(\lambda)}\int  d\mu(\phi,\bar \phi)\frac{4 \bar \phi \phi }{1+ 4\lambda \bar \phi \phi}  e^{  - \frac{1}{2}\log (1-4 \lambda \bar \phi \phi)  } \\
&=& 1 -  4\lambda + \cdots .
\eea
We recover the familiar logarithmic form of the action and resolvent 
of the  intermediate field theory. However our LVR representation 
is not the intermediate field representation. Indeed in the LVR representation 
the argument of the log is quadratic in complex fields similar to the initial fields
although it would be linear in the single real field $\sigma$ of the intermediate field representation.
We should rather think to the fields of the LVR as to what remains of the initial fields  \emph{after having forced 
integration of one particular marked $\bar \phi$ field per vertex}.

\subsection{The $(\bar \phi \phi)^3$ theory}
\medskip
In this case $g= (-\lambda)^{\frac{1}{3}} = e^{i\pi/3 } \lambda^{1/3}$,  
$\lambda = - g^3$ and $z = g^3 (\bar \phi \phi)^2  = - \lambda (\bar \phi \phi)^2 $. 
Equation \eqref{gencatalan} is now
\bee
zT_3^3(x) -T_3(z) +1 =0 . \label{cardano}
\ee
which is soluble by radicals. Introducing 
\bee u:= - \frac{27 z}{4} = -\frac{27 }{4}g^3 (\bar \phi \phi)^2  = \frac{27 }{4}\lambda (\bar \phi \phi)^2, \ee 
Cardano's solution is
\bee
T_3(z) =\frac{\Delta_+(u)   - \Delta_{-}(u)}{\sqrt{-3z}} = 1  +z   +  3z^2  + \cdots,
\ee
where
\bee\Delta_\pm (u) :=  
\biggl(\sqrt{1 +u} \pm \sqrt u   \biggr)^{1/3} = 1  \pm  \frac{1}{3}  \sqrt {u} + \frac{u}{18}  \mp  \frac{4 u^{3/2}}{81} - \frac{35u^2}{1944} + \cdots .
\ee
Defining $h (u) := \frac{1}{ \sqrt{1 +u}}$, 
we can compute the derivatives 
\bea \Delta'_\pm =\frac{d}{du}\Delta_\pm (u) &=&  \frac{1}{6} \biggl((1  + u)^{-1/2} \pm  u^{-1/2} \biggr) \biggl(\sqrt{1  + u} \pm \sqrt {u}   \biggr)^{-2/3}
\nonumber\\
&=& \pm \frac{1}{6\sqrt {u(1+u)}}  \Delta_\pm (u) =  \pm \frac{h }{6\sqrt {u}}  \Delta_\pm (u) \label{derdelta}.
\eea
Hence
\bee
zT_3'(z) =\frac{27 \sqrt{-z}}{4 \sqrt 3}[\Delta'_+(u) - \Delta'_{-}(u)] -  \frac{1}{2\sqrt{-3z}}[\Delta_+ (u) - \Delta_{-}(u)] .
\ee
\eqref{f0} gives
\bea  F_{3} &=& \sum_n   {{3n}\choose{n}} z^n = 2z T_3' +T_3 = h \frac{ \Delta_+ + \Delta_{-} }{2} ,\\
  S_{3}  &=& \log F_3 =   -\frac{1}{2} \log (1+u)   + \log  \frac{ \Delta_+ + \Delta_{-} }{2}  
\eea
The $u$ derivatives of $S_3$ give access to its $g$ derivatives since $u= -\frac{27 }{4}g^3 (\bar \phi \phi)^2 $,
hence
\bee  \frac{\partial u}{\partial g}  =  -\frac{81 }{4}g^2 (\bar \phi \phi)^2 .
\ee
For instance
\bea
\frac{\partial S_3}{\partial g}  &=& \frac{81 g^2 (\bar \phi \phi)^2}{4} \biggl(\frac {1}{2(1 + u)}   -\frac{ \Delta'_+  + \Delta'_-}{ \Delta_+  + \Delta_-}\biggr)
\nonumber\\
&=&  \frac{81 g^2 (\bar \phi \phi)^2}{8} \biggl(h^2  - \frac{h }{3\sqrt {u}} \frac{  \Delta_+  - \Delta_-}{ \Delta_+  + \Delta_-}\biggr).
\eea 

We remark that the quotient $\frac{  \Delta_+  - \Delta_-}{ \Delta_+  + \Delta_-}= \frac{A-B}{A+B}$ for $A =  \Delta_+ $, $B= \Delta_- $ simplifies,
using that $(A+B)(A^2 - AB + B^2) = A^3 + B^3$ and $(A-B)(A^2 - AB + B^2) = A^3 - B^3 -2AB(A-B)$. Remarking that 
in our case $AB = \Delta_+ \Delta_-= 1$ we find
\bea \frac{  \Delta_+  - \Delta_-}{ \Delta_+  + \Delta_-} &=& h [\sqrt u   -  ( \Delta_+  - \Delta_-)] \label{quotient}\\
\frac{\partial S_3}{\partial g}  &=&\nonumber \frac{81 g^2 (\bar \phi \phi)^2}{8} \biggl(h^2  
- \frac{h^2 }{3\sqrt {u}}     [\sqrt u   -  ( \Delta_+  - \Delta_-)] \biggr)\\
 &=&\nonumber \frac{81 g^2 (\bar \phi \phi)^2}{8} \biggl[\frac {2}{3} h^2 
+ \frac{h^2 }{3\sqrt {u}} ( \Delta_+  - \Delta_-) \biggr]\\
 &=& \frac{27 g^2 (\bar \phi \phi)^2}{4 (1+u)} 
\biggl[ 1  +  \frac{ \Delta_+  - \Delta_- } {2 \sqrt u} \biggr]
\label{vertexder}
\eea
from which we find
\bea  G^c_{3,1} (\lambda) &=&\nonumber 1 + \frac{g}{Z_3(\lambda)}\int  d\mu(\phi ,\bar \phi) \frac{\partial S_3 }{\partial g}  e^{S_3}\\
&=&\nonumber  1 -   \frac{27 \lambda}{4Z_3(\lambda)}\int  d\mu(\phi ,\bar \phi)  \frac{(\bar \phi \phi)^2}{1+u} \biggl[ 1  +  \frac{ \Delta_+  - \Delta_- } {2 \sqrt u} \biggr]  e^{S_3}\\
&=&  1 -  18 \lambda  +\cdots .
\eea

\subsection{The $(\bar \phi \phi)^4$ theory}
\medskip

In this case $g= (-\lambda)^{\frac{1}{4}} = e^{i\pi/4 } \lambda^{1/4}$,  $\lambda = - g^4$ and $z = g^4 (\bar \phi \phi)^3  = - \lambda (\bar \phi \phi)^3 $. 
Equation \eqref{gencatalan} is now
\bee
zT_4^4(z) -T_4(z) +1 =0 . \label{cardano}
\ee
which is still soluble by radicals.
Denoting
\bee
v= \frac{z^{1/3}}{ 2^{1/3} } \Big{[} \Big{(}1+\sqrt{ 1 - \frac{2^8}{3^3} z  } \Big{)}^{1/3} +
\Big{(} 1 -\sqrt{1 - \frac{2^8}{3^3} z} \Big{)}^{1/3} \Big{]},
\ee
then
\bee \label{cat4}
T_4(z)= \frac{  (1+4v)^{1/4} - \bigl[2-(1+4v)^{1/2} \bigr]^{1/2}  }{2 (vz)^{1/4}} \;.
\ee

We can then compute
\bea  F_{4} &=& \nonumber\frac{T_4}{4 - 3T_4} = 3z T_4' +T_4  \\
&=&  \frac{  (1+4v)^{1/4} - \bigl[2-(1+4v)^{1/2} \bigr]^{1/2}  }{8 (vz)^{1/4}  - 3 \bigl[ (1+4v)^{1/4} - \bigl[2-(1+4v)^{1/2} \bigr]^{1/2}\bigr] }
\eea
from which $S_4 = \log F_4$ can be derived explicitly.

\end{document}